\documentclass{article}
\usepackage{authblk}
\usepackage[british]{babel}
\usepackage{csquotes}

\author[1]{Andreani Petrou\footnote{andreani.petrou@oist.jp}}
\author[1]{Shinobu Hikami\footnote{hikami@oist.jp}}
\affil[1]{Okinawa Institude of Science and Technology Graduate University, 

1919-1 Tancha, Okinawa 904-0495, Japan.}

\usepackage{fancyhdr} 
\usepackage[a4paper, left=20mm, right=20mm, top=20mm, bottom=2.5cm, footskip=1.2cm]{geometry}  

\usepackage{braket}

\title{Harer--Zagier formulas for families of twisted hyperbolic knots}

\usepackage{mathtools}
\usepackage{amssymb}
\usepackage{tikz}
\usetikzlibrary{spath3, intersections, hobby}

\usepackage{xfrac}

\usepackage[utf8]{inputenc}
\usetikzlibrary{shapes,snakes}
\usepackage{graphicx} 
\usepackage{subcaption}

\usepackage{braids}
\usetikzlibrary{decorations.markings,hobby,knots,celtic}

\usepackage{graphicx}

\newcommand*{\img}[1]{%
    \raisebox{-.3\baselineskip}{%
        \includegraphics[
        height=\baselineskip,
        width=\baselineskip,
        keepaspectratio,
        ]{#1}%
    }%
}

\usepackage{wrapfig}
\graphicspath{ {./images/} }

\begin{document}
\date{}
\maketitle 

\vspace{0.1cm}

\begin{abstract}
    In an attempt to generalise knot matrix models  for  non-torus knots, which currently remains an open problem, we derived formulas for the Harer--Zagier transform of the HOMFLY--PT polynomial  for some infinite families of twisted hyperbolic knots. Among them, we found a family of Pretzel knots for which the transform has a fully factorised form, while for the remaining families considered it consists of sums of factorised terms.  Their zeros have a remarkable structure as the modulus of their product in all cases equals unity. 
\end{abstract}


\section{Introduction}

Among the many uses of knots by humans since antiquity,  their ability to store information is remarkable. In particular, the ancient Chinese and Incan civilisations used knotted strings as an alternative to writing. In the late 19th century, Lord Kelvin, through his vortex atom hypothesis \cite{kelvin1867vortex}, envisioned to use knots to encode information about nature. Although his theory thwarted, it gave birth to Knot theory, the mathematical study of knots. Among its major achievements was the discovery of knot polynomial invariants, such as the Alexander and Jones polynomials, or their 2 variable generalisation, called the HOMFLY-PT polynomial. About a century later, a revolutionary work by E. Witten \cite{witten1989quantum} attributed a physical interpretation to such invariants, as observables  of Chern--Simons theory, hence reconnecting knots with physics and resulting into a fruitful interchange. Indeed, more recently with the development of matrix models, there has been an active effort to explore this interrelation more deeply; and it is towards this goal that the present work aims to contribute.

\subsection{Chern--Simons theory and knot invariants}\label{sec:CS}
Chern--Simons (CS) theory is a  Topological Quantum Field Theory on a 3-dimensional manifold that is invariant under the action of a gauge group $G$. 
The Wilson loop operators  $W_\mathcal{K}^R$
are the traces of holonomies around a knot $\mathcal{K}$, evaluated in an irreducible representation $R$ of $G$. The averages of these $\langle W^R_\mathcal{K}\rangle$ are quantum, gauge invariant observables of the theory. 
The special case when the manifold is $\mathbb{S}^{3}$,  $G=SU(N)$ and $R=\square$ (the fundamental representation) the observables yield the    HOMFLY-PT  polynomial of the knot $\mathcal{K}$ (defined below in sec. \ref{sec:HOMFLYandHZ}) as  $\bar{H}_{\mathcal{K}}(q^{N},q)=\langle W_\mathcal{K}^{\square} \rangle$,
where $q$ depends on $N$ and $k$,
 the level (or coupling constant) of CS theory \cite{witten1989quantum,1995marinolabastida}. It can be \enquote{colored} by different choices of  irreducible representations $R$, resulting in the colored HOMFLY (henceforth omitting --PT) polynomial $\bar{H}_{\mathcal{K}}^R(q^{N},q)=\langle W_\mathcal{K}^{R} \rangle$.
It is a generalisation of both the Jones and  Alexander polynomials, which correspond to the particular cases $N=2$ and $N=0$, respectively. 

CS theory on $\mathbf{S}^3$ with gauge group $U(N)$ also admits a matrix model formulation with measure for the average  (up to constant factors) given by
\begin{equation}\label{eq:CSmatrixmodel}
\langle F \rangle_{CS}\sim\intop F \prod_{i<j}^{N}\left(2\sinh\left(\frac{x_{i}-x_{j}}{2}\right)\right)^{2}\prod_{i=1}^{N}dx_{i}e^{-x_{i}^{2}/2g},
\end{equation}
where $\{x_i\}_{i=1}^{N}$ are the eigenvalues of an $N\times N$ Hermitian matrix, $F$ is a function of the $\{x_i\}$, $g=\frac{2\pi i}{k+N}$ and the factor in the bracket is known as the trigonometric Van-der-Monde function \cite{tierz2004soft}. 

\subsection{Knot matrix models}\label{sec: KnotMM}
More recently, Morozov et al. \cite{morozov2021harer} conjectured a connection between knot polynomial invariants and matrix models via the \emph{superintegrability condition }
\begin{equation}\label{eq:superintegrability}\langle\chi^R\rangle_\mathcal{K}=\Bar{H}^R_\mathcal{K}(q^N,q).
\end{equation}
Superintegrability means that a complete set of averages are explicitly calculable; and it is established that  for (Hermitian) eigenvalue matrix models the averages of characters are known to be again characters, i.e $\langle\chi^R\rangle \sim \chi^R$ \cite{ITOYAMA_2012,Mironov_2021}.
Due to the dependence of Wilson loop averages on representations, 
knot polynomial invariants can be thought of as  non-trivial generalisations of characters\footnote{In particular, the HOMFLY  polynomial for torus knots can be expressed in terms of Schur functions, see \cite{2013barkowski} for details.}, hence allowing to use the condition  $ \left\langle character \right\rangle=knot\;polynomial$ as the defining property (\ref{eq:superintegrability}).

Knot matrix models are, thus far, only  consistently defined  for the particular case of torus knots\footnote{A torus knot (or link, when $(m,n)$ are not coprime)  is described algebraically as the intersection of the 3-sphere with  a singular complex curve $V=\{(\alpha,\beta)\in\mathbb{C}^{2}| \alpha^{m}-\beta^{n}=0\}$, i.e. $T(m,n)=T(n,m)=V\cap\mathbb{S}^{3}$. The  integers $(m,n)$ give the number of strands (toroidal windings)  and number of leaves (poloidal windings), respectively.} $T(m,n)$,
for which there exists 
  an eigenvalue matrix model, the TBEM model \cite{tierz2004soft,Brini_2012}, providing an explicit measure in the left hand side of (\ref{eq:superintegrability}), given by
\begin{equation}\label{eq:torusknotmatrixmodel}
  \langle \chi^{R}\rangle_{T(m,n)}\sim \intop \chi^{R} \prod_{i<j}^{N}\sinh\left(\frac{x_{i}-x_{j}}{m}\right)\sinh\left(\frac{x_{i}-x_{j}}{n}\right)\prod_{i=1}^{N}dx_{i}e^{-x_{i}^{2}/2g}.    
  \end{equation} 
Here $q=e^{\frac{g}{2mn}}$ and note that 
the trigonometric Van-der-Monde function is $(m,n)$-deformed, but otherwise this expression is  identical with the one for the CS matrix model (\ref{eq:CSmatrixmodel}).
  
The \textbf{ Harer--Zagier (HZ) transform}, which  is a discrete version of the Laplace transform  in $N$
explicitly given by
\begin{equation}
    Z_{\mathcal{K}}(q,\lambda)=\sum_{N=0}^{\infty}\Bar{H}_{\mathcal{K}}(q^N,q)\lambda^{N}
    \label{harerzagier}
\end{equation}
provides an alternative manifestation of superintegrability:
\begin{equation}\label{factorizability}
     the\; HZ\; tranforms\; are\; completely\; factorised\; rational\; functions,
\end{equation}
i.e. they have zeroes and poles at positive and negative powers of $q$. This is true for the case of torus knots, as shown in \cite{morozov2021harer}  using the  quantum groups technology and reconfirmed (via a different method) in the present work; and it should be a minimum consistency requirement of any extension of the definition (\ref{eq:superintegrability}) of knot matrix models to other families of knots. As a first check,
the HZ formula for the HOMFLY polynomial of  the simplest hyperbolic  knot,  the figure-8, was also computed in \cite{morozov2021harer}. This turned out  to be not factorisable, and hence superintegrability fails in this case.

However, continuing this effort, in this article we derive the HZ formulas for some infinite families of \enquote{twisted} hyperbolic knots (which shall be described in more detail below) and examine their factorisability properties (sec.  \ref{sec:twistedhyp}), their $q\rightarrow1$ expansion (sec. \ref{sec:expansion}), their poles (sec. \ref{sec:poles}) and zero loci (sec. \ref{sec:zerolocus}). The Appendix includes the HZ formulas and the  $q\rightarrow1$ expansion coefficients for some further families of twisted hyperbolic knots.

\section{The HOMFLY polynomial and its Harer--Zagier tranform}\label{sec:HOMFLYandHZ}
\begin{wrapfigure}{R}{0.22\textwidth}
\centering
\includegraphics[width=0.25\textwidth]{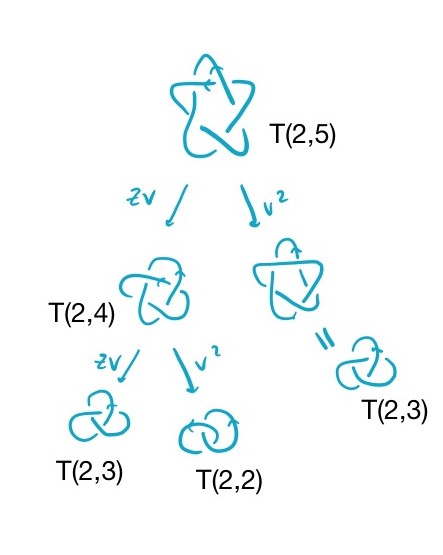}
\caption{Resolving tree for $T(2,5)$}\label{fig:skeintree}
\end{wrapfigure}
The  HOMFLY polynomial $H_{\mathcal{K}}(v,z)$ of an oriented knot is a Laurent polynomial in two variables, defined by the normalisation condition $H_{\text{unknot}}=1$ and the \emph{skein relation} 
\begin{equation}\label{eq:skein}
    v^{-1}H_{L_{+}}(v,z)-vH_{L_{-}}(v,z)=zH_{L_{0}}(v,z)
\end{equation}
 where $L_{+}=$\img{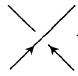},
 $L_{-}=$\img{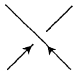},
 $L_{0}=$\img{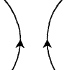}.
For two disconnected knots $\mathcal{K}_1$ and $\mathcal{K}_2$ the product formula $H_{\mathcal{K}_1	\sqcup \mathcal{K}_2}=(v^{-1}-v)z^{-1}H_{\mathcal{K}_1}H_{\mathcal{K}_2}$ holds \cite{murasugi1996knot}.
  
 The unnormalised HOMFLY $\bar{H}_{\mathcal{K}}(v,z)$, is obtained by multiplying with an overall  factor  $-(v^{-1}-v)z^{-1}=:\bar{H}_{unknot}(v,z)$; and with the substitution $v=q^{N}$ and $z=q-q^{-1}$, where $q=e^{\sfrac{\pi i}{(k+N)}}$, we obtain $\bar{H}_{\mathcal{K}}(q^{N},q)$ as arises from CS theory described above\footnote{Due to a discrepancy in conventions between the mathematics and physics literature, this holds up to some minus signs. For instance, an extra overall minus sign is included in $\bar{H}_{unknot}$  in order to be in agreement with   the Wilson loop average of the circle in standard framing, as derived in \cite{witten1989quantum}. Such ambiguities, however,  do not affect the essence of the results in this article.}. 
The HOMFLY polynomial can be combinatorially computed using skein trees, as the one shown in the figure \ref{fig:skeintree} for the example of the torus knot $T(2,5)$, with the help of which
we have derived  recursive or explicit formulas and computed their HZ transform for the following  families of  knots.

\subsection{Torus knots and links}
The HOMFLY polynomial for 2--stranded torus knots and links\footnote{Whenever we refer to torus links, we assume that all the components have parallel orientation, as  shown for example for $T(2,4)$ and $T(2,2)$ in fig. \ref{fig:skeintree}. The recursive formula (\ref{eq:recursive(2,n)links}) is not valid for links with different relative orientation.} can be obtained by the following recursive relations with initial condition $H_{T(2,2)}=\frac{v}{z} \left(1-v^2+z^2\right)$
\begin{equation}\label{eq:recursive(2,n)links}
    H_{T(2,n)}=v^{2}H_{T(2,n-2)}+zvH_{T(2,n-1)},\; \forall\; n\geq3.
\end{equation}
For odd $n=2k+1$ with $k=1,2,3,...$ this can be restricted to knots only:
\begin{equation}\label{eq:recursive(2,n)knots only}
    H_{T(2,2k+1)}=v^{2}H_{T(2,2k-1)}+z^{2}\sum_{j=1}^{k}v^{2j}H_{T(2,2(k-j)+1)}+v^{2k}(1-v^{2})H_{T(2,1)}.
\end{equation}
For 3--stranded torus knots and links there are 3 different recursive formulas corresponding to $n\mod3=\{0,1,2\}$, with initial condition $H_{T(3,3)}=v^4 z^2 \left(2-v^2+z^2\right)+v^4z^{-2} \left(1-v^2+z^2\right) \left(1-v^2+2 z^2\right)$.\\
- $\forall\;n\mod3=2$, i.e. $n=2,5,8,...$
\begin{equation}\label{eq:recursive(3,n)links}
    H_{T(3,n)}=v^{2}H_{T(3,n-1)}+z^{2}\sum_{j=1}^{n-1}v^{2j}H_{T(3,n-j)}+v^{2(n-1)}(1-v^{2})H_{T(3,1)},
\end{equation}
- $\;\forall\;n\mod3=1 \;(\geq4)$, i.e. $n=4,7,10,...$
\begin{equation}\label{eq:recursive(3,n)links1mod3}
   H_{T(3,n)}=v^{4}H_{T(3,n-2)}+v^{2}z^{2}H_{T(3,n-1)}+2z^{2}\sum_{j=2}^{n-1}v^{2j}H_{T(3,n-j)}+2v^{2(n-1)}(1-v^{2})H_{T(3,1)},
\end{equation}
- $\forall\;n\mod3=0\;(\geq6)$, i.e. $n=6,9,12,...$
\begin{equation}\label{eq:recursive3n0mod3}
H_{T(3,n)}=v^{6}H_{T(3,n-3)}+v^{2}z^{2}H_{T(3,n-1)}+2v^{4}z^{2}H_{T(3,n-2)}+3z^{2}\sum_{j=3}^{n-1}v^{2j}H_{T(3,n-j)}+3v^{2(n-1)}(1-v^{2})H_{T(3,1)},
\end{equation}
where $T(m,1)$ is the unknot $\forall\; m$.
To our knowledge, formulas (\ref{eq:recursive(2,n)knots only})--(\ref{eq:recursive3n0mod3}) are new results as they are nowhere to be found in the literature. Due to the fact that skein trees are not unique and grow fast even at $m=4$, we were unable to obtain a  recursive relation for general\footnote{However, we did obtain the general recursion formula $V_{T(m,n)}(q)=q^{2(m-1)}V_{T(m,n-2)}(q)+(1-q^{2(m-1)})q^{(m+1)(n+1)}$ for the single variable Jones polynomial corresponding to $N=2$, i.e. $V_{\mathcal{K}}(q)=H_{\mathcal{K}}(q^2,q-q^{-1})$.} $(m,n)$. Hence, we used instead the explicit formula for the HOMFLY polynomial of torus knots only (i.e. for $m,n$ coprime) given in \cite{giasemidis2014torus}, which in our conventions reads 
\begin{equation*}
    \Bar{H}_{T(m,n)}(q^{N},q)=\frac{q^N-q^{-N}}{q-q^{-1}}(q^{N}q)^{(m-1)(n-1)}\frac{1-q^{-2}}{1-q^{-2m}}{\displaystyle \sum_{\beta=0}^{m-1}q^{-2n\beta}\left(\prod_{i=1}^{\beta}\frac{q^{2N}q^{2i}-1}{q^{2i}-1}\right)\left(\prod_{j=1}^{m-1-\beta}\frac{q^{2N}-q^{2j}}{1-q^{2j}}\right)}.
\end{equation*}
The corresponding HZ transform can be computed by  applying the geometric series to each $q^{N}$ power. Doing this calculation for sufficiently many torus knots of fixed $m$ and arbitrary $n$, we inductively deduced the following factorised formula
 \begin{equation}
     Z_{T(m,n)}(q,\lambda)=\frac{\lambda\prod_{j=0}^{m-2}\left(1-\lambda q^{(m+1)n+m-2-2j}\right)}{\prod_{j=0}^{m}\left(1-\lambda q^{(m-1)n+m-2j}\right)}.
 \end{equation}
Under $q\rightarrow q^{-1}$ and $\lambda\rightarrow q^{mn}\lambda$  (the latter making the $m\leftrightarrow n$ symmetry of torus knots more explicit), this indeed
reproduces the result of \cite{morozov2021harer}, as claimed in the introduction.

 \subsection{Twisted hyperbolic knots}\label{sec:twistedhyp}
 
\begin{figure}[!h]
\begin{subfigure}[h]{0.24\linewidth}
\includegraphics[width=\textwidth]{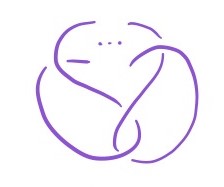}
\caption{$\overline{2k}\;\overline{2}$}\label{2k 2}
\end{subfigure}
\hfill
\begin{subfigure}[h]{0.24\linewidth}
\includegraphics[width=0.9\textwidth]{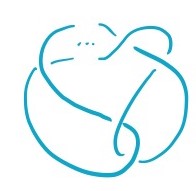}
\caption{$\overline{2k+1}\;\overline{2}$}\label{2k+1 2}
\end{subfigure}%
\hfill
\begin{subfigure}[h]{0.24\linewidth}
\includegraphics[width=0.87\textwidth]{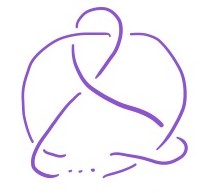}
\caption{$\overline{2k+1}\;\overline{1}\;\overline{2}$}\label{2k+1 1 2}
\end{subfigure}%
\begin{subfigure}[h]{0.24\linewidth}
\includegraphics[width=0.9\textwidth]{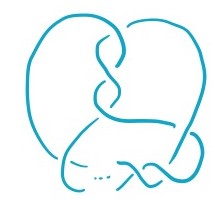}
\caption{$(2k+2)\;3$}\label{2k+2 3}
\end{subfigure}%
\caption{Some families of twisted hyperbolic knots. }\label{fig:twistedhyp1}
\end{figure}

We repeated the above calculation for some  families of \enquote{twisted}\footnote{Note the difference with the standard knot theory jargon, in which \emph{twist knots} all have unkotting number $1$ (as e.g. the families $\overline{2k}\;\overline{2}$, $\overline{2k+1}\;\overline{2}$); and they don't restrict to hyperbolic as they also include 2-stranded  torus knots $T(2,n)$.} hyperbolic knots, obtained  as follows. Given a projection of a simple knot, which can be thought of as a \emph{generating knot}, we choose a point where two strands are parallel (adjacent) to each other  and cut it open there, as in fig. \ref{2k 2}--\ref{2k+2 3}. Introducing a number of whole twists at the place indicated with 3 dots in the figure, and then reconnecting the strands,  yields an infinite family of knots with only even or odd number of crossings. The reason we avoid half twists is because they sometimes result into more than one component, i.e. a link, which we would like to omit in the current treatment. The twisted hyperbolic families are labelled using Conway notation\footnote{If the reader is unfamiliar with Conway notation, we refer to Chapter 2 of \cite{adams1994knot} for a concise and comprehensive introduction.}, in which the juxtaposed numbers indicate the number of crossings of the individual tangles used to compose the knot, while an over-line (instead of the standard minus sign) is used to denote negative tangles. The total number of crossings $n$ of the knot is equal to the sum of its Conway numbers. Successive members of a family correspond to increasing $k=1,2,3,...$, each having $k-1$ additional whole twists, which can be thought of as $2k-1$ extra \enquote{bubbles}.
The generating knot, corresponding to $k=1$, is chosen in a way such that the bubbles consist of positive crossings (c.f. $L_{+}$ in (\ref{eq:skein}))  which amounts to sometimes using the mirror of the knots listed in the Rolfsen table \cite{adams1994knot} (while we shall not be careful to explicitly mention this whenever using Rolfsen notation, it should be clarified by the respective Conway notation).
 Four such families, shown in fig. \ref{fig:twistedhyp1},   along with the obtained results for their HOMFLY polynomials and HZ transforms, are given below. Some further examples are included in the Appendix.
 \\
(a) The family $\overline{2k}\;\overline{2}$  is generated by the figure-8 knot $\overline{2}\;\overline{2}$, or $4_1$, and includes the knots $6_1,\;8_1,\;10_1$, in Rolfsen notation. Its (unnormalised) HOMFLY polynomial is $\Bar{H}_{\overline{2k}\;\overline{2}}(v,z)=\frac{v-v^{-1}}{z}(v^{2k}(1-v^{-2})+v^{-2}-z^{2}\sum_{j=0}^{k-1}v^{2j}$), while its Harer--Zagier transform  in terms of the total number of crossings $n=2+2k$ is
   \begin{multline*}
    Z_{\overline{2k}\;\overline{2}}(q,\lambda)=\frac{\lambda\left(1+\lambda q^{-5}\right)\left(1-\lambda^{2}q^{3n-8}\right)}{\left(1-\lambda q^{-1}\right)\left(1-\lambda q^{-3}\right)\left(1-\lambda q^{n-5}\right)\left(1-\lambda q^{n-3}\right)\left(1-\lambda q^{n-1}\right)}\\
    -\frac{\lambda^{2}q^{n-3}\left(\left(q^{2}+1+q^{-2}\right)\left(1-\lambda q^{n-7}\right)+q^{-3}\left(q^{n-3}+q^{-n+3}\right)\left(1-\lambda q^{n-1}\right)-q^{n}\left(1-\lambda q^{-n-7}\right)\right)}{\left(1-\lambda q^{-1}\right)\left(1-\lambda q^{-3}\right)\left(1-\lambda q^{n-5}\right)\left(1-\lambda q^{n-3}\right)\left(1-\lambda q^{n-1}\right)}
\end{multline*}\\
(b) The family $\overline{2k+1}\;\overline{2}$ 
is generated by $\overline{3}\;\overline{2}$ or $5_2$; includes the knots $\;7_2$ and $9_2$; \\
$\Bar{H}_{\overline{2k+1}\;\overline{2}}(v,z)=\frac{v-v^{-1}}{z}(v^{2(k+1)}(1-v^{2})+v^{2}+z^{2}\sum_{j=1}^{k+1}v^{2j})$; $n=3+2k$
 \begin{equation*}
    Z_{\overline{2k+1}\;\overline{2}}(q,\lambda)=\frac{\lambda\left(\left(1-\lambda q^{3}\right)\left(1-\lambda q^{n-2}\right)\left(1-\lambda q^{2n+3}\right)-\lambda\left(1-\lambda q^{n+2}\right)\left(q^{n}-q^{5}\right)\left(1-q^{n-3}\right)\right)}{\left(1-\lambda q\right)\left(1-\lambda q^{3}\right)\left(1-\lambda q^{n-2}\right)\left(1-\lambda q^{n}\right)\left(1-\lambda q^{n+2}\right)}
\end{equation*}\\
(c) The family $\overline{2k+1}\;\overline{1}\;\overline{2}$ 
is generated by $\overline{3}\;\overline{1}\;\overline{2}$ or $6_2$; includes $\;8_2$ and $10_2$; \\
$\Bar{H}_{\overline{2k+1}\;\overline{1}\;\overline{2}}(v,z)=v^{-2}\Bar{H}_{T(2,2k+1)}(v,z)-zv^{-1}\Bar{H}_{T(2,2k+2)}(v,z)$; $n=4+2k$
\begin{equation*}
Z_{\overline{2k+1}\;\overline{1}\;\overline{2}}(q,\lambda)=\frac{\lambda\left(\left(1+\lambda q^{n-9}\right)\left(1+\lambda q^{3n-7}\right)-\lambda q^{2n-8}\left(q^{-2}+q^{2}\right)\left(q^{-n+3}+q^{n-3}\right)\right)}{\left(1-\lambda q^{n-7}\right)\left(1-\lambda q^{n-5}\right)\left(1-\lambda q^{n-3}\right)\left(1-\lambda q^{n-1}\right)}
\end{equation*}\\
(d) The family $(2k+2)\;3$ 
is generated by $7_3$, but can also  be thought of as being generated by the $2\;3$ projection of $5_2$, corresponding to $k=0$; includes $9_3$;  $\Bar{H}_{(2k+2)\;3}(v,z)=v^{2}\Bar{H}_{T(2,2k+3)}(v,z)+zv\Bar{H}_{T(2,2k+2)}(v,z)$; $n=5+2k$
\begin{equation*}
    Z_{(2k+2)\;3}(q,\lambda)=\frac{\lambda\left(\left(1-\lambda q^{n-2}\right)\left(1-\lambda q^{3n-2}\right)-\lambda q^{2n-2}\left(q-q^{-1}\right)\left(q^{n-5}-q^{-n+5}\right)\right)}{\left(1-\lambda q^{n-4}\right)\left(1-\lambda q^{n-2}\right)\left(1-\lambda q^{n}\right)\left(1-\lambda q^{n+2}\right)}
\end{equation*}

We deduce that the HZ transforms for these families of twisted hyperbolic knots still have completely factorised denominators but the numerators now consist of sums of two or more factorised terms. It is worth pointing out the similarity of the recursive formulas in the latter two cases with the one for 2-strand torus knots (\ref{eq:recursive(2,n)links}), hence resulting in almost factorised HZ functions. The only exception among these families is the case $5_2$ (i.e. $\overline{3}\;\overline{2}$ or $2\;3$), which has a completely factorised HZ function
\begin{equation}\label{eq:Z5_2}
    Z_{5_2}(q,\lambda)=\frac{\lambda\left(1-\lambda q^{13}\right)}{\left(1-\lambda q\right)\left(1-\lambda q^{5}\right)\left(1-\lambda q^{7}\right)}.
\end{equation}
Beyond these families, we computed the HZ transform for the HOMFLY polynomial of all knots in the Rolfsen table with up to 8 crossings\footnote{We have also considered composite knots $\mathcal{K}_{1}\#\mathcal{K}_{2}$, for which $H_{\mathcal{K}_{1}\#\mathcal{K}_{2}}=H_{\mathcal{K}_{1}}H_{\mathcal{K}_{2}}$, but they seem to not have a factorised HZ transform even when $\mathcal{K}_{1,2}$ are both torus knots.}. Among them we found that, apart from $5_2$, $8_{20}$ also has an HZ transform with a completely factorised form. Subsequently, we realised that there is a whole family of twisted hyperbolic knots generated by a 6 crossing projection of $5_2$.
\begin{figure}[!h]
   \centering
   \includegraphics[scale=0.9]{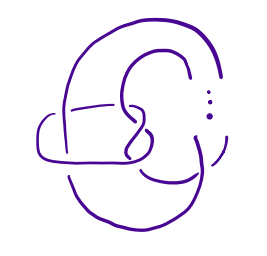}
\caption{$P(\overline{2},3,\overline{2k+1})$}
   \label{fig:P(2,3,2k+1)}
\end{figure}
\setlength{\textfloatsep}{1pt}\\
 These are the Pretzel knots $P(\overline{2},3,\overline{2k+1})$, shown in fig. \ref{fig:P(2,3,2k+1)}, in which $5_2$ corresponds to $k=0$, while it includes the knots $8_{20}$ at $k=1$ and $10_{125}$ at $k=2$. Their HOMFLY polynomial and the corresponding HZ transforms are 
\begin{equation*}
\Bar{H}_{P(\overline{2},3,\overline{2k+1})}=v^{-2}\Bar{H}_{P(\overline{2},3,\overline{2k-1})}+z^{2}\sum_{j=1}^{k}v^{-2j}\Bar{H}_{P(\overline{2},3,\overline{2(k-j)+1})}-v^{-2k}(1-v^{2}+z^{2})\Bar{H}_{T(2,3)},
\end{equation*}
\begin{equation}
    Z_{P(\overline{2},3,\overline{2k+1})}(q,\lambda)=\frac{\lambda \left(1-\lambda q^{13-2 k}\right) \left(1-\lambda q^{3 (1-2 k)}\right)}{\left(1-\lambda q^{1-2 k}\right) \left(1-\lambda q^{3-2 k}\right) \left(1-\lambda q^{5-2 k}\right) \left(1-\lambda q^{7-2 k}\right)}
\end{equation}
which agrees with eq. (\ref{eq:Z5_2}) at $k=0$. From this expression it is clear that the family $P(\overline{2},3,\overline{2k+1})$ satisfies the property (\ref{factorizability}) and hence it might be possible to derive an explicit measure for the average  $\langle...\rangle_{P(\overline{2},3,\overline{2k+1})}$, which would give the first working definition of a knot matrix model for hyperbolic knots. This will be the subject of future investigation.\\

\section{Analysis of HZ functions}
A few remarks about the results listed in the previous section are in order. \\
\textbf{Remark 1} At $q=1$, all HZ formulas reduce to $Z_{\mathcal{K}}(1,\lambda)=\frac{\lambda}{(1-\lambda)^2}$. Moreover, in the limits $q\rightarrow\infty$ and $q\rightarrow0$, only the formulas $Z_{\overline{2k+1}\;\overline{2}}$, $Z_{(2k+2)\;3}$ (corresponding to knots with odd number of crossings $n$) and $Z_{\overline{2k+1}\;\overline{1}\;\overline{2}}$ for $k\geq3$ (or $n\geq10$) have a finite values, equal to $1/\lambda$ and $\lambda$, respectively. These coincide with the hyperbolic families with a non-factorised HZ transform that have no zeros on the negative real axis (c.f. sec. \ref{sec:zerolocus} below).\\
\textbf{Remark 2} If $\lambda$ is set to $q$ or $q^{-1}$, the HZ transform of some twisted hyperbolic knots becomes factorised. Examples are $Z_{\overline{5}\;\overline{2}}(q,\lambda=q)=q(1-q^{16})/((1-q^{2})(1-q^{6})(1-q^{10}))$, $Z_{4\;3}(q,\lambda=q^{-1})=- (1+ q^{10})/(q(1-q^2)(1-q^6))$ and $Z_{6\;3}(q,\lambda=q)=- q(1+ q^{14})/((1-q^6)(1-q^{10}))$.\\
\textbf{Remark 3} All of the above formulas are invariant under $q\mapsto q^{-1}$ and $\lambda\mapsto \lambda^{-1}$, i.e. $Z_{\mathcal{K}}(q,\lambda)=Z_{\mathcal{K}}(q^{-1},\lambda^{-1})$, while the modular transformations $q\mapsto -q^{-1}$ and $\lambda\mapsto -\lambda^{-1}$ yield $Z_{\mathcal{K}}\mapsto -Z_{\mathcal{K}}$.

\subsection{Expansion for $q$ close to $1$}\label{sec:expansion}
The limit $q\rightarrow1$ is equivalent to the limit of large $k$, which is referred to as the weak coupling limit in the physics literature \cite{witten1989quantum}. In this regime, fixing $\lambda=1$,  we can set $q=e^x$ for  $|x|\ll1$ and expand the HZ formulas in powers of $x$. The expansions always take the form $Z_{\mathcal{K}}(e^{x},1)= \sum_{i=-1}^{\infty} a_{2i}^{\mathcal{K}}x^{2i}$, where $a_{2i}^{\mathcal{K}}\in \mathbb{Q}$ have denominators that are multiples of a fixed odd number.  
We have explicitly computed $a_{-2}^{\mathcal{K}}$ for the above twisted families of knots
\begin{equation*}
    a_{-2}^{\overline{2k}\;\overline{2}}=-\frac{1}{3}+\frac{4}{(n-5) (n-3) (n-1)},\;\;a_{-2}^{\overline{2k+1}\;\overline{2}}=\frac{1}{3}+\frac{4}{(n-2) n (n+2)},
\end{equation*}
\begin{equation*}
a_{-2}^{\overline{2k+1}\;\overline{1}\;\overline{2}}=\frac{3 \left(n^2-6 n+13\right)}{(n-7) (n-5) (n-3) (n-1)},\;\;a_{-2}^{(2k+2)\;3}=\frac{3 \left(n^2-4 n+8\right)}{ (n-4) (n-2) n (n+2)},
\end{equation*}
\begin{equation*}
a_{-2}^{P(\overline{2},3,\overline{2k+1})}=\frac{3 (n-19)}{(n-13) (n-11) (n-9)}.
\end{equation*}

\subsection{Poles and holomorphicity}\label{sec:poles}
As can be easily seen from the fully factorised denominators of the HZ formulas  their $\lambda$ poles  lie at positive and negative powers of $q$, hence they  lie on the unit circle (recall $q=e^{\pi i/(k+N)})$, while there is an additional pole at $\lambda=\infty$.
The sum of the residues  over all the finite $\lambda$ poles equals $1$. In fact, it is interesting to note that this can be deduced by considering just the first (factorised) part of the HZ formulas, as the sum of the residues of the $\lambda^2$ term always vanishes. Finally, adding the residue at the pole at infinity, which always equals $-1$, the total sum becomes $0$. 
Moreover, at fixed $\lambda=1$, the $q$-poles of the HZ formulas lie at $0,\;1,\infty$ and at roots of unity. Again the sum of all the residues, including infinity, equals $0$.
Via Cauchy theorem, this implies that the HZ formulas are holomorphic in the extended complex $\lambda$ and $q$ planes.

\subsection{Zero locus}\label{sec:zerolocus}
\begin{figure}[!h]
\begin{subfigure}[h]{0.3\linewidth}
\includegraphics[width=0.8\textwidth]{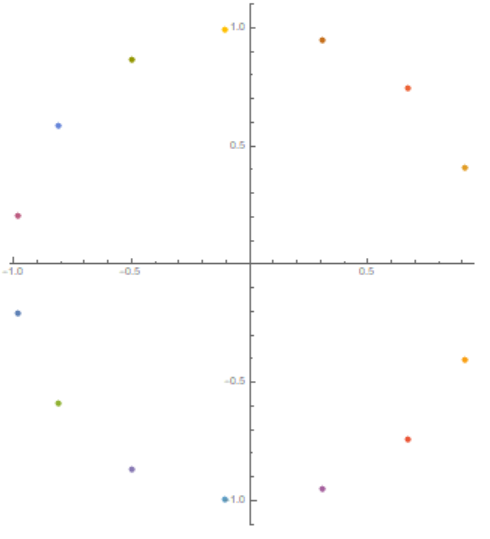}
\caption{$T(2,5)$}
\end{subfigure}
\hfill
\begin{subfigure}[h]{0.3\linewidth}
\includegraphics[width=0.8\textwidth]{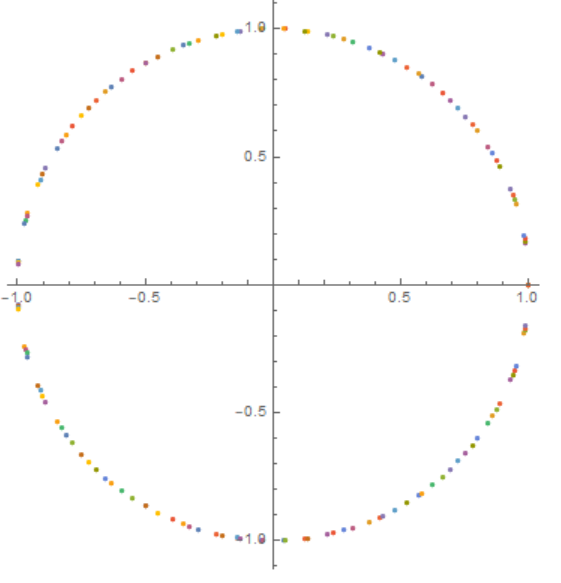}
\caption{$T(5,6)$}
\end{subfigure}%
\hfill
\begin{subfigure}[h]{0.3\linewidth}
\includegraphics[width=0.8\textwidth]{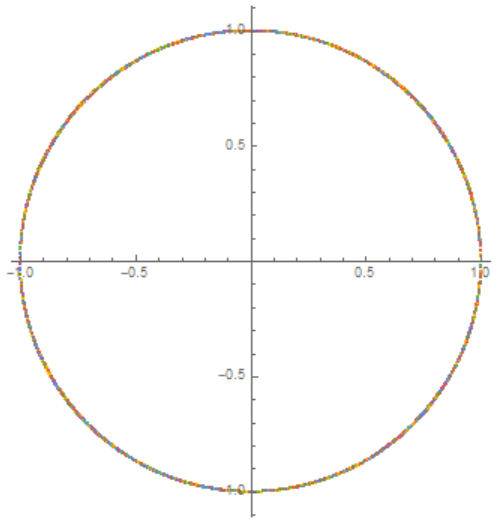}
\caption{$T(15,17)$}
\end{subfigure}%
\caption{For torus knots $T(m,n)$ all zeros  have norm equal to $1$, i.e. they lie on the unit circle. As $(m,n)$ increase these become more dense, but none seems to lie on the real axis. }\label{fig:zerostorus}
\end{figure}

\begin{figure}[!h]
\begin{subfigure}[h]{0.3\linewidth}
\includegraphics[width=0.8\textwidth]{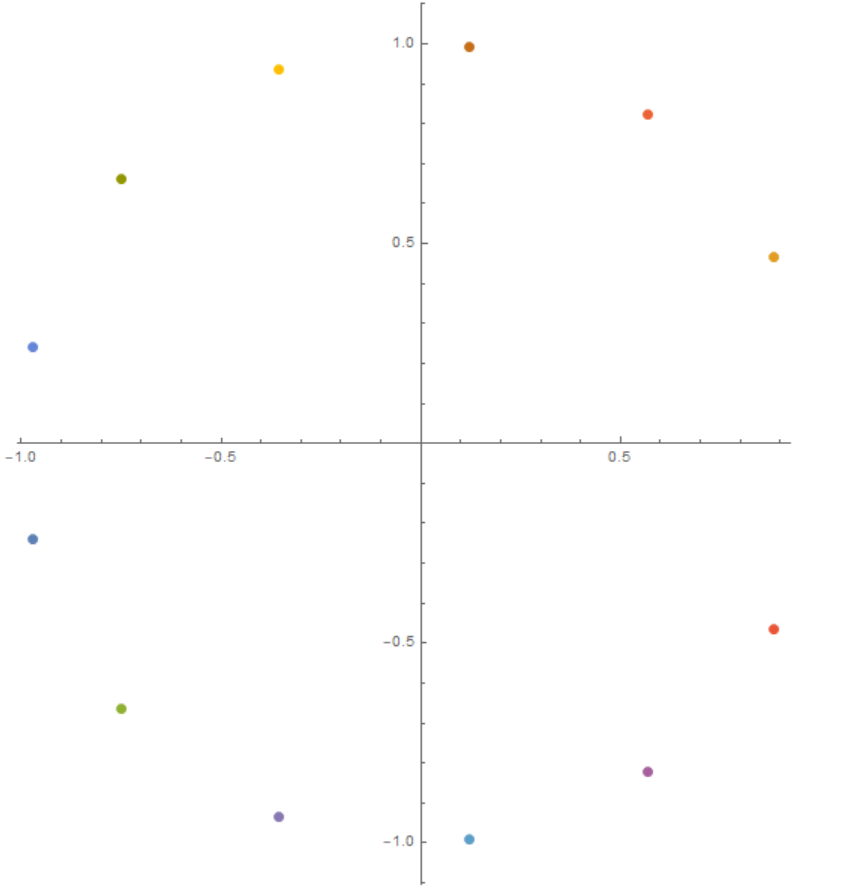}
\caption{$P(\overline{2},3,\overline{1})$ or $5_2$}
\end{subfigure}
\hfill
\begin{subfigure}[h]{0.3\linewidth}
\includegraphics[width=0.8\textwidth]{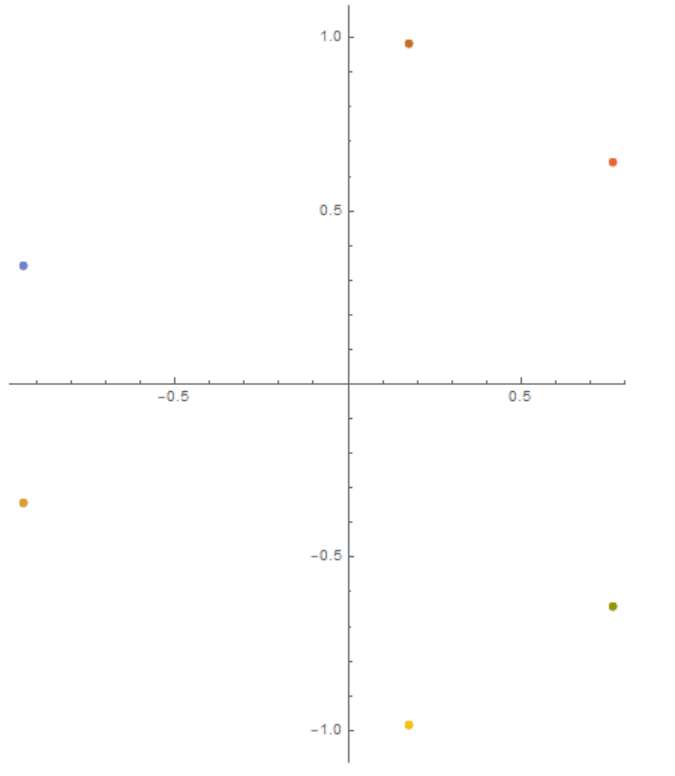}
\caption{$P(\overline{2},3,\overline{5})$ or $10_{125}$}
\end{subfigure}%
\hfill
\begin{subfigure}[h]{0.3\linewidth}
\includegraphics[width=0.8\textwidth]{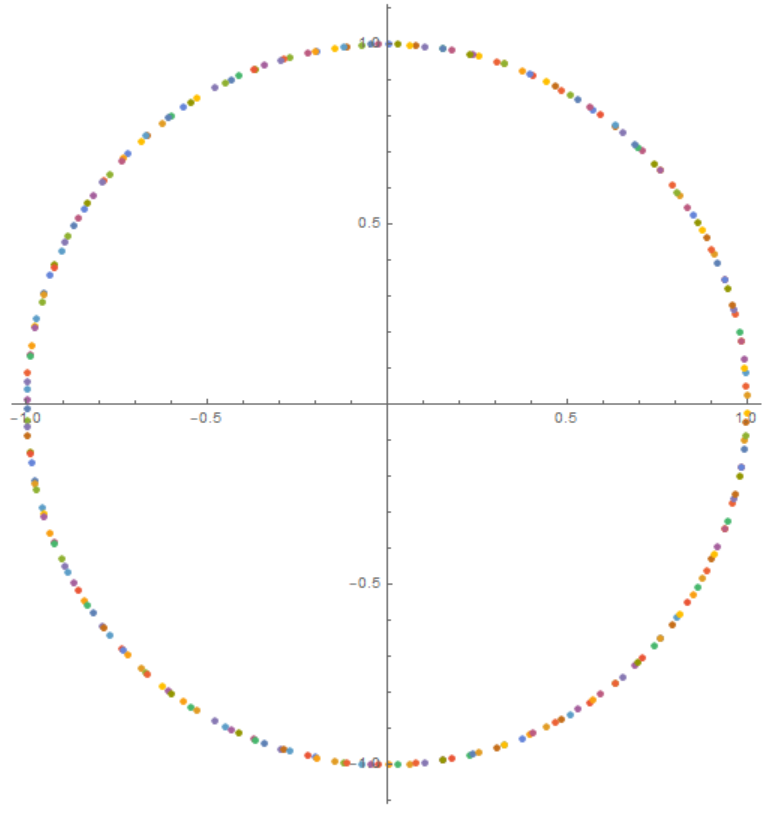}
\caption{$P(\overline{2},3,\overline{85})$}
\end{subfigure}%
\caption{For the Pretzel family $P(\overline{2},3,\overline{2k+1})$ all zeros  have norm equal to $1$, i.e. they lie on the unit circle. For $k=1,2$ their density descreases but for $k\geq3$ it increases. Again, none of the zeros lies on the real axis. }\label{fig:zerospretzel}
\end{figure}

\begin{figure}[!h]
\begin{subfigure}[h]{0.3\linewidth}
\includegraphics[width=\textwidth]{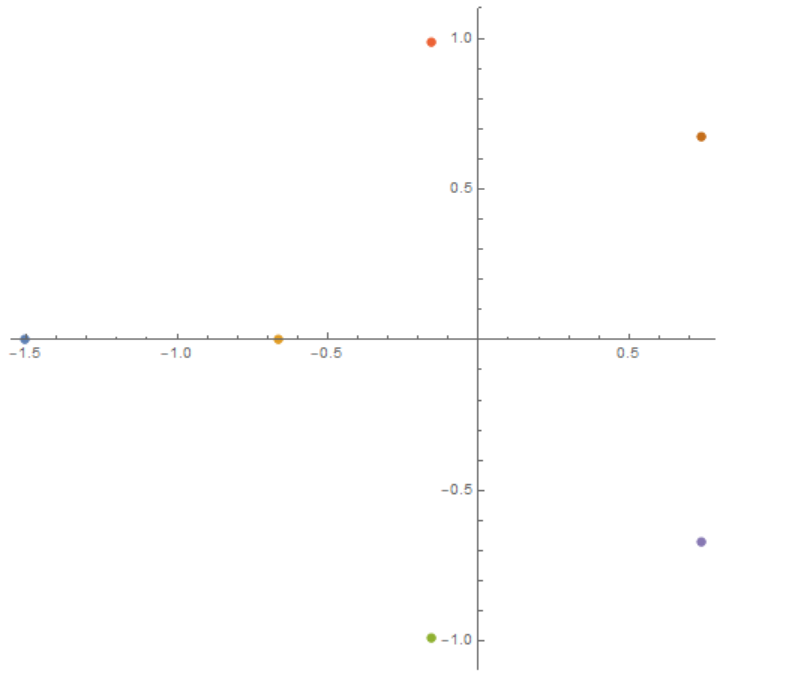}
\caption{$\overline{2}\;\overline{2}$ or $4_1$}
\end{subfigure}
\hfill
\begin{subfigure}[h]{0.3\linewidth}
\includegraphics[width=\linewidth]{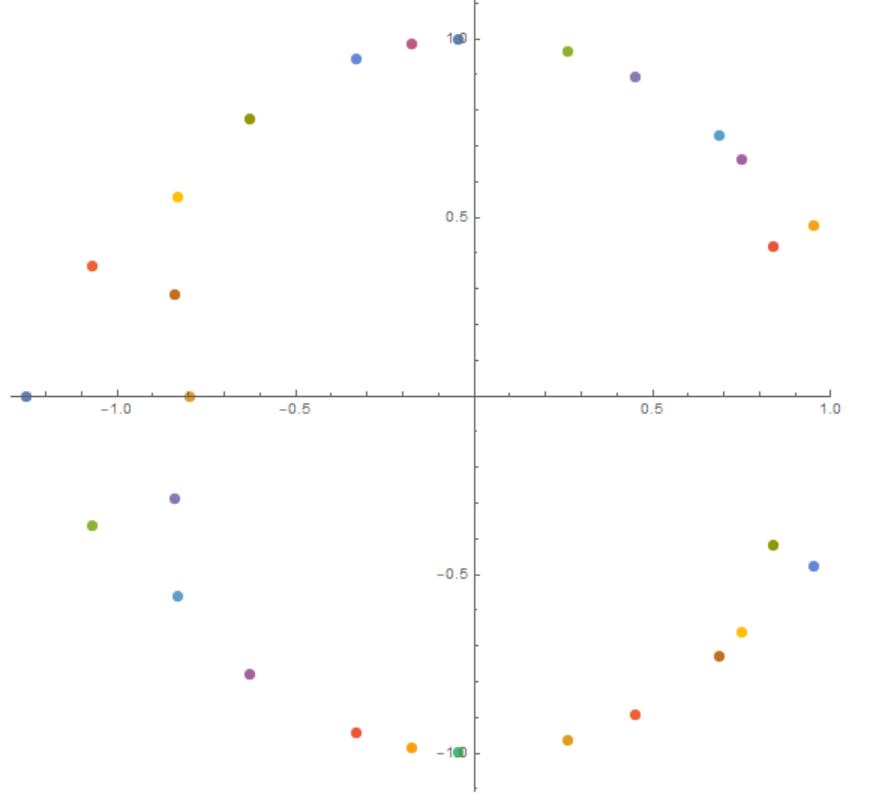}
\caption{$\overline{10}\;\overline{2}$}
\end{subfigure}%
\hfill
\begin{subfigure}[h]{0.3\linewidth}
\includegraphics[width=\textwidth]{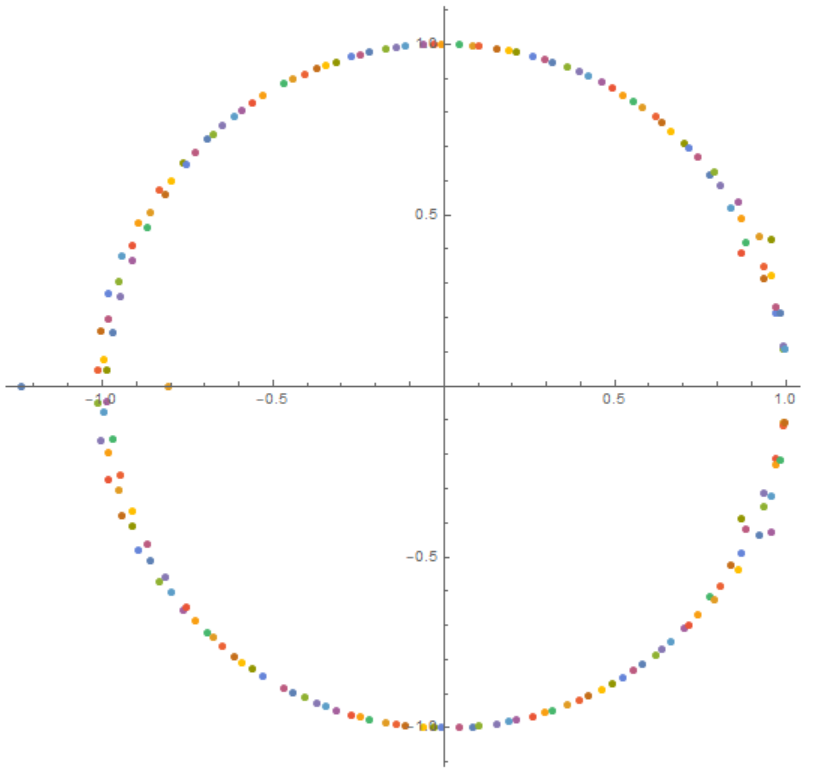}
\caption{$\overline{58}\;\overline{2}$}
\end{subfigure}%
\caption{For hyperbolic knots in the $\overline{2k}\;\overline{2}$ family, most zeros still lie on the unit circle, with exception a finite number of pairs which, when multiplied have norm equal to $1$, i.e. they are related by conformal inversion. One such pair lies on the negative real axis $\forall\;n$.}\label{fig:zerosn1}
\end{figure}

\begin{figure}[!h]
\begin{subfigure}[h]{0.3\linewidth}
\includegraphics[width=\textwidth]{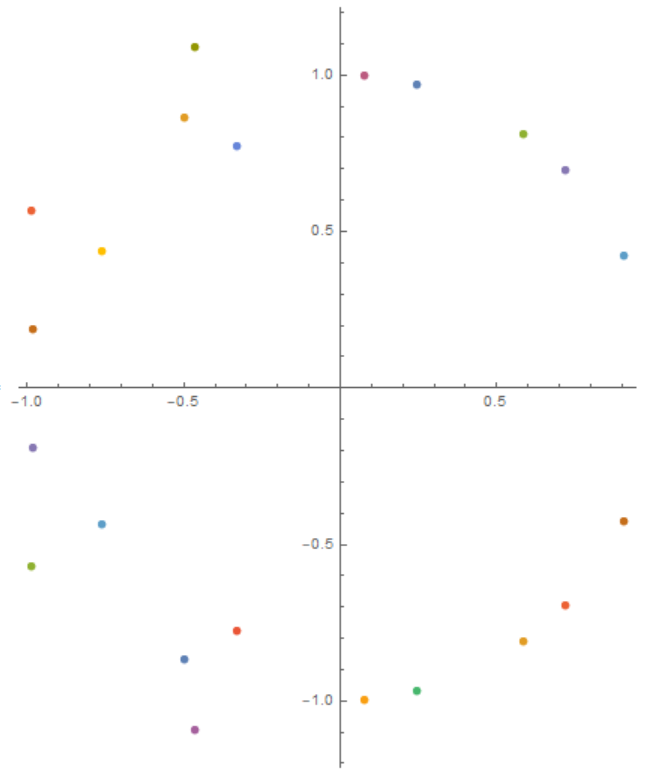}
\caption{$\overline{5}\;\overline{2}$ or $7_2$}
\end{subfigure}%
\hfill
\begin{subfigure}[h]{0.3\linewidth}
\includegraphics[width=\textwidth]{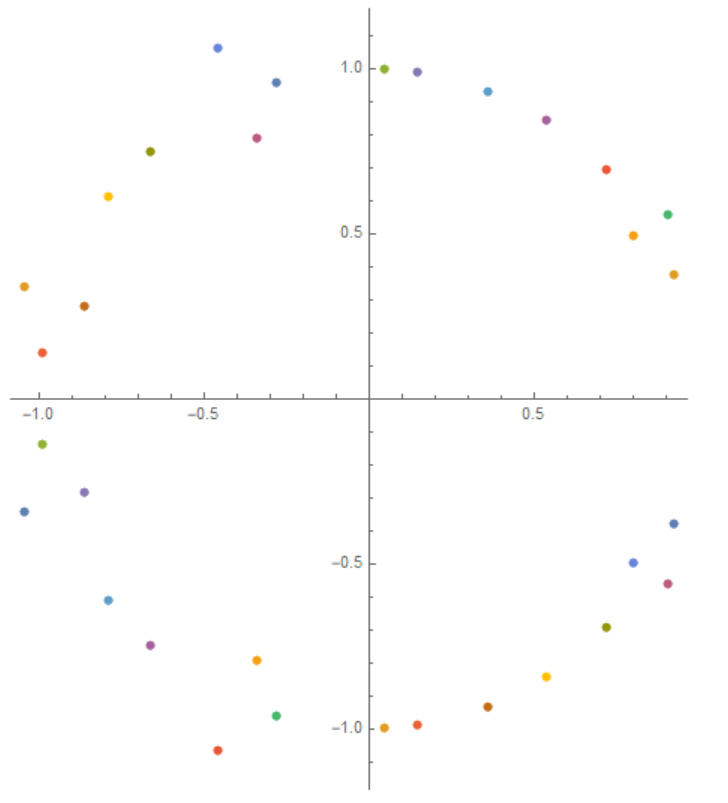}
\caption{$\overline{9}\;\overline{2}$}
\end{subfigure}
\hfill
\begin{subfigure}[h]{0.3\linewidth}
\includegraphics[width=\textwidth]{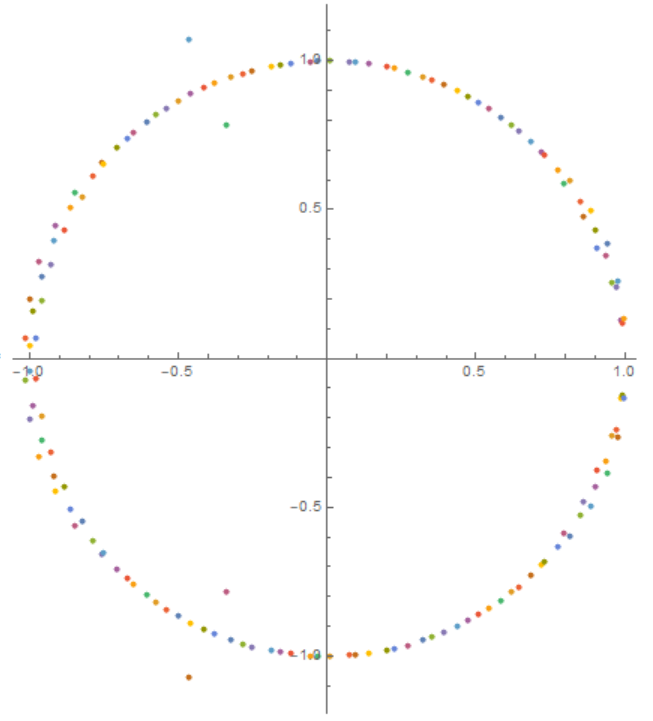}
\caption{$\overline{47}\;\overline{2}$}
\end{subfigure}%
\caption{For hyperbolic knots in the $\overline{2k+1}\;\overline{2}$ family, the zero structure for $n=3+2k\geq 7$ shows again a deviation from the unit  circle only in conformal pairs. As for torus knots, none of them lies on the real axis.}\label{fig:zerosn2odd}
\end{figure}

\begin{figure}[!h]
\begin{subfigure}[h]{0.3\linewidth}
\includegraphics[width=\textwidth]{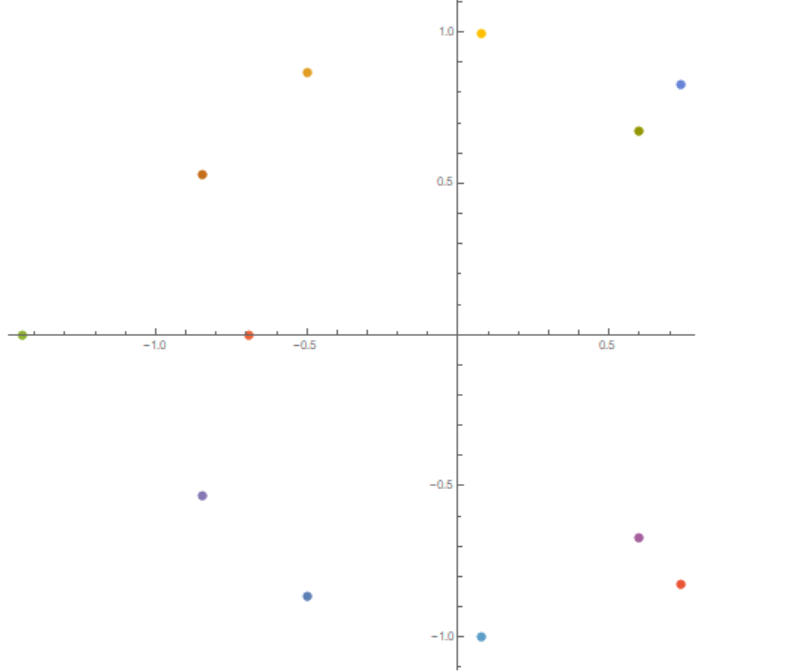}
\caption{$\overline{3}\;\overline{1}\;\overline{2}$ or $6_2$}
\end{subfigure}
\hfill
\begin{subfigure}[h]{0.3\linewidth}
\includegraphics[width=\textwidth]{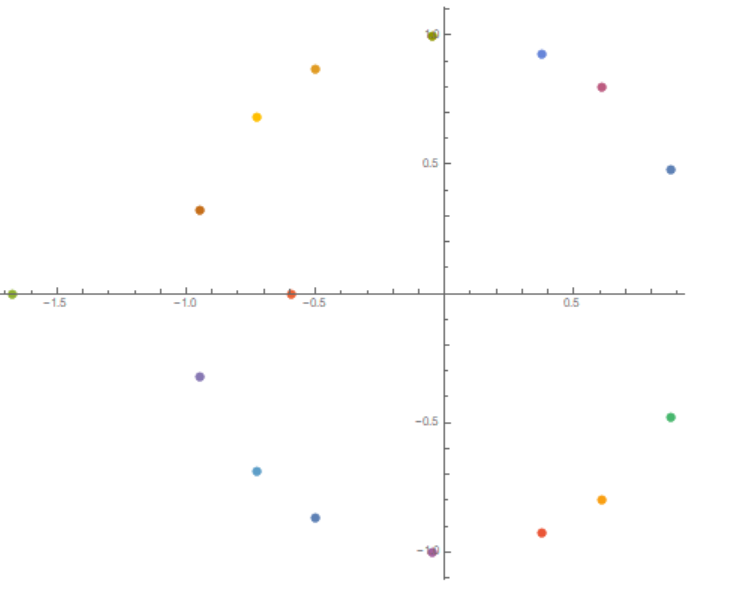}
\caption{$\overline{5}\;\overline{1}\;\overline{2}$ or $8_2$}
\end{subfigure}%
\hfill
\begin{subfigure}[h]{0.3\linewidth}
\includegraphics[width=0.8\textwidth]{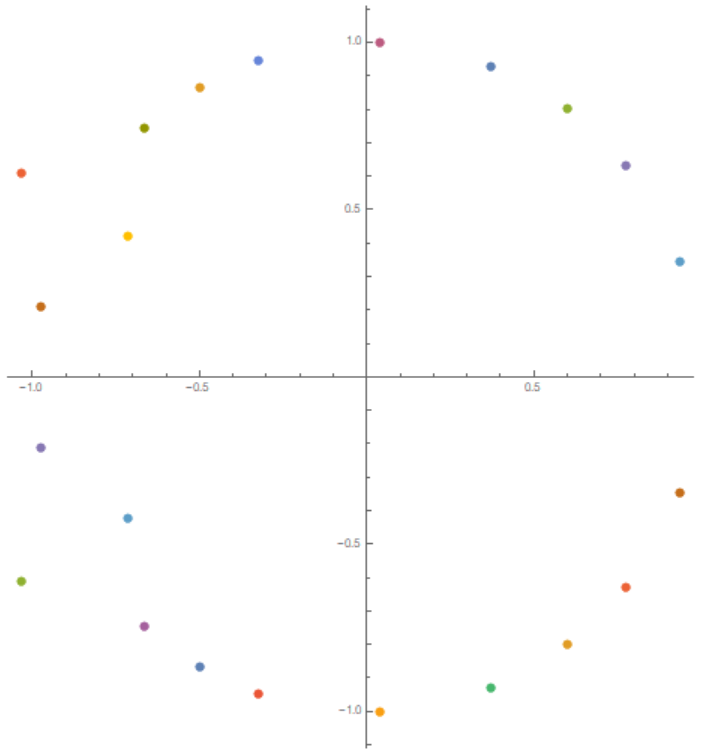}
\caption{$\overline{7}\;\overline{1}\;\overline{2}$ or $10_2$}
\end{subfigure}%
\caption{Zero locus for some hyperbolic knots in the $\overline{2k+1}\;\overline{1}\;\overline{2}$ family. For $n=6,8$ there is a pair of zeros on the real axis, as in the $\overline{2k}\;\overline{2}$ family (c.f. fig. \ref{fig:zerosn1}), but this stops being the case  $\forall\; n\geq 10$.}\label{fig:zerosn2even}
\end{figure}

It is also of interest to consider the zeros of the above derived HZ formulas. In the figures \ref{fig:zerostorus}--\ref{fig:zerosn2even} below we plot the vanishing sets $\{q\in\mathbb{C}|Z_{\mathcal{K}}(q,1)=0\}$ with fixed $\lambda=1$, for a few examples of both  torus and twisted hyperbolic knots.
 From these plots we deduce that when the  HZ formulas are factorised, as it is the case for $P(\overline{2},3,\overline{2k+1})$ and torus knots, the zeros have unit norm, i.e. they lie on the unit circle. When the HZ transform consists of sums of factorised terms, as for the majority of the twisted hyperbolic knots considered, deviations from the circle arise in conformal pairs, i.e. there are zeros of the form $ae^{i\phi}$ and $\frac{1}{a}e^{i\phi}$ with $|a|\neq1$.
 The plots for the $(2k+2)\;3$ family have similar traits as the ones in fig. \ref{fig:zerosn2odd} for $\overline{2k+1}\;\overline{2}$, and hence  are omitted. The resemblance of these results with the zeros of the characteristic function for the exponents of a singular complex curve studied in  \cite{saito1983zeroes} is astounding. Moreover, there might be a relation of these zero structures to the zeros of the Riemann-$\zeta$ function \cite{Brzin2017RandomMT}. In fact, it is remarkable that such zero structures appear in various areas of mathematics, but they lack a systematic study.  Hence,  a more in depth exploration of these connections deserves to be the subject of future research.

\paragraph{Acknowledgements} 

A.P. is grateful to Roland van der Veen for his feedback and suggestions at the early stages of this project. We would also like to thank Reiko Toriumi for participating in our discussions. This work is supported by JSPS Kakenhi 19H01813.

\clearpage



\bibliographystyle{unsrt}
 \typeout{} 
\bibliography{reference}

\clearpage

\section*{Appendix A}

Here we include the computed HOMFLY polynomials and their HZ transforms for some further families of twisted hyperbolic knots. The coefficient of the poles in the $q\rightarrow1$ expansion of sec. \ref{sec:expansion} are also given.\\ 
 \begin{figure}[!h]
\begin{subfigure}[h]{0.24\linewidth}
\includegraphics[width=\textwidth]{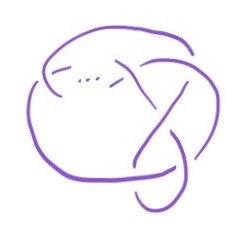}
\caption{$2k\;1\;1\;2$}
\end{subfigure}
\hfill
\begin{subfigure}[h]{0.24\linewidth}
\includegraphics[width=0.87\textwidth]{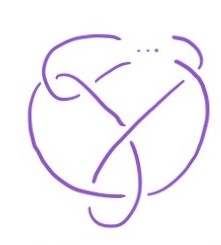}
\caption{$2\;(2k-1)\;1\;2$}
\end{subfigure}%
\hfill
\begin{subfigure}[h]{0.24\linewidth}
\includegraphics[width=0.73\textwidth]{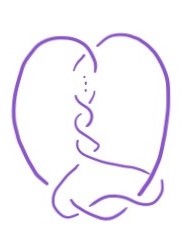}
\caption{$4\;(2k+2)$}
\end{subfigure}%
\begin{subfigure}[h]{0.24\linewidth}
\includegraphics[width=0.9\textwidth]{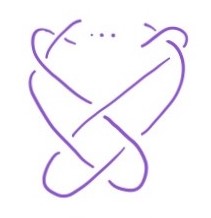}
\caption{$(2k+2)\;1\;3$}
\end{subfigure}%
\end{figure}\\
(a) $\Bar{H}_{2k\;1\;1\;2}(v,z)=v^{-2}(1+z^{2})\Bar{H}_{T(2,2k+1)}(v,z)-zv^{-3}\Bar{H}_{T(2,2k)}(v,z)$; includes  $6_{3},8_{7},10_{5}$; $n=4+2k$ 
\begin{equation*}
Z_{2k\;1\;1\;2}(q,\lambda)=\frac{\lambda\left(\left(1-\lambda q^{n-13}\right)\left(1-\lambda q^{3n-11}\right)-\lambda q^{2n-12}\left(q-q^{-1}\right)\left(q^{2}+q^{-2}\right)\left(q^{-n+4}-q^{n-4}\right)\right)}{\left(1-\lambda q^{n-9}\right)\left(1-\lambda q^{n-7}\right)\left(1-\lambda q^{n-5}\right)\left(1-\lambda q^{n-3}\right)}
\end{equation*}
\begin{equation*}
   a_{-2}^{2k\;1\;1\;2}= \frac{3 \left(n^2-14 n+37\right)}{(n-9) (n-7) (n-5) (n-3)}
\end{equation*}
(b) $\Bar{H}_{2\;(2k-1)\;1\;2}(v,z)=v^{-2}(1+z^{2})\Bar{H}_{\overline{2k-1}\;\overline{2}}(v,z)-zv^{-3}\Bar{H}_{T(2,2)}(v,z)$; includes $6_{3},8_{8},10_{34}$; $n=4+2k$
 \begin{multline*}
    Z_{2\;(2k-1)\;1\;2}(q,\lambda)=\lambda\{\left(1-\lambda q^{-7}\right)\left(1-\lambda q\right)\left(1-\lambda q^{n-7}\right)\left(1-\lambda q^{2n-5}\right)\\
+\lambda q^{n-2}\left(1-\lambda q^{-7}\right)\left(\left(q^{-n+5}+q^{n-5}\right)\left(1-\lambda q^{n-7}\right)+q^{-1}\left(1+\lambda q^{3}\right)\left(-1+q^{n-8}\right)\right)\\
+\lambda q^{-5}\left(1-\lambda q\right)\left(\left(1-\lambda q^{3n-13}\right)\left(1-q^{2}\right)^{2}-q^{n}\left(1+\lambda q\right)\left(1-q^{n-10}\right)\right)\\
-\lambda q^{2n-11}\left(q^{2}+q^{-2}\right)\left(1-\lambda q\right)\left(1-\lambda q^{-n+3}\right)\}\\
\times\left(\left(1-\lambda q^{-3}\right)\left(1-\lambda q^{-1}\right)\left(1-\lambda q\right)\left(1-\lambda q^{n-7}\right)\left(1-\lambda q^{n-5}\right)\left(1-\lambda q^{n-3}\right)\right)^{-1}
\end{multline*}
\begin{equation*}
    a_{-2}^{2\;(2k-1)\;1\;2}=-\frac{7}{3}+\frac{4}{(n-7)(n-5)(n-3)}
\end{equation*}
(c) $\Bar{H}_{{4\;(2k+2)}}(v,z)=v^{-2}\Bar{H}_{\overline{2(k+1)}\;\overline{2}}(v,z)-zv^{2k-1}\Bar{H}_{T(2,2)}(v,z)-z(v-v^{-1})\sum_{j=0}^{k-1}v^{2j}$; includes $8_3$, $10_3$; $n=6+2k$ 
 \begin{multline*}
    Z_{4\;(2k+2)}(q_{,}\lambda)=\lambda\{\left(1+\lambda q^{-9}\right)\left(1-\lambda q^{n-7}\right)\left(1+\lambda q^{2n-7}\right)\left(1+\lambda^{2}q^{n-10}\right)\\
-\lambda q^{n-7}\left(1+\lambda q^{-9}\right)\left(q^{2}\left(1-\lambda q^{-5}\right)\left(1+\lambda q^{2n-9}\right)+q^{-2}\left(1-q^{n-2}\right)\left(1+\lambda^{2}q^{n-4}\right)\right)\\
-\lambda q^{n-3}\left(1-\lambda q^{-9}\right)\left(1+q^{n-10}\right)\left(1+\lambda^{2}q^{n-8}\right)\\
-\lambda q^{-3}\left(2\left(1+\lambda q^{-7}\right)\left(1-\lambda^{2}q^{4n-20}\right)+q^{2n-14}\left(1-\lambda q^{-1}\right)^{2}\left(1-\lambda q^{3}\right)\right)\\
+2\lambda^{2}q^{-8}\left(\left(1-\lambda q^{n-3}\right)\left(1+q^{3n-14}\right)+q^{n+1}\left(q+q^{-1}\right)\left(1-\lambda q^{2n-19}\right)\right)\}\\
\times\left(\left(1-\lambda q^{-5}\right)\left(1-\lambda q^{-3}\right)\left(1-\lambda q^{-1}\right)\left(1-\lambda q^{n-9}\right)\left(1-\lambda q^{n-7}\right)\left(1-\lambda q^{n-5}\right)\left(1-\lambda q^{n-3}\right)\right)^{-1}
\end{multline*}
\begin{equation*}
   a_{-2}^{4\;(2k+2)}= \frac{1}{15}-\frac{8 (n-6)}{(n-9) (n-7) (n-5) (n-3)}
\end{equation*}
(d) $\Bar{H}_{(2k+2)\;1\;3}(v,z)=v^{-2}(1+z^{2})\Bar{H}_{\overline{2(k+1)}\;\overline{2}}(v,z)-zv^{2k-3}\Bar{H}_{T(2,2)}(v,z)-z(v-v^{-1})\sum_{j=-1}^{k-2}v^{2j}$;\\ includes $8_4$, $10_4$; $n=6+2k$
 \begin{multline*}
    Z_{(2k+2)\;1\;3}(q_{,}\lambda)=\frac{\lambda(\left(1+\lambda q^{-11}\right)\left(1-\lambda q^{n-5}\right)\left(1+\lambda q^{2n-13}\right))}{\left(1-\lambda q^{-3}\right)\left(1-\lambda q^{-5}\right)\left(1-\lambda q^{n-9}\right)\left(1-\lambda q^{n-7}\right)\left(1-\lambda q^{n-5}\right)}\\
   +\frac{\lambda^{2}q^{-7}\left(\left(1-q^{n}\right)\left(1-q^{n-2}\right)(1-\lambda q^{n-13})-q^{n-5}\left(q^{-2}+q^{2}\right)\left(q^{n-5}+q^{-n+5}\right)\left(1-\lambda q^{n-5}\right)\right)}{\left(1-\lambda q^{-3}\right)\left(1-\lambda q^{-5}\right)\left(1-\lambda q^{n-9}\right)\left(1-\lambda q^{n-7}\right)\left(1-\lambda q^{n-5}\right)}
\end{multline*}
\begin{equation*}
   a_{-2}^{(2k+2)\;1\;3}= \frac{1}{15}-\frac{8}{(n-9) (n-7) (n-5)}
\end{equation*}

\end{document}